\documentclass[11pt]{article}

\usepackage[utf8]{inputenc}
\usepackage[T1]{fontenc}
\usepackage{amsmath,amssymb}
\usepackage{graphicx}
\usepackage{booktabs}
\usepackage{multirow}
\usepackage{subcaption}
\usepackage{array}
\usepackage{xcolor}
\usepackage[hidelinks]{hyperref}
\usepackage[margin=1in]{geometry}
\usepackage{authblk}

\newcommand{\dn}{$\downarrow$}
\newcommand{\up}{$\uparrow$}

\title{Beyond Generative Decoding: Discriminative Hidden-State Readout
from a Native Omni-Modal LLM for Multimodal Sentiment Analysis}

\author[1]{Bin Wen}
\author[1]{Tien-Ping Tan\thanks{Corresponding author: Tien-Ping Tan
(e-mail: \texttt{tienping@usm.my}).}}
\affil[1]{\small School of Computer Sciences, Universiti Sains Malaysia,
Penang, Malaysia. \quad Email: \texttt{wenbin@student.usm.my},
\texttt{tienping@usm.my}}
\date{May 2026}

\begin{document}
\maketitle

\begin{abstract}
Multimodal sentiment analysis (MSA) seeks to infer human affect from
language, acoustic, and visual signals. Recent work increasingly builds on
large language models (LLMs) and large multimodal models (LMMs), which are
typically queried in a generative manner: the model is prompted to emit the
sentiment score as a text string that is then parsed into a number. This
generative readout is convenient, but it ties a continuous regression
target to discrete autoregressive decoding, and its costs for MSA have
rarely been measured. We revisit the readout mechanism for LMM-based MSA and
propose a discriminative formulation built on the Thinker module of a native
omni-modal LLM (Qwen2.5-Omni-7B). Instead of decoding a number, we take the
hidden state of the last non-padding token at the final layer and map it to
a continuous sentiment score with a lightweight regression head, in a single
forward pass. The backbone is adapted with 4-bit quantization and low-rank
adaptation (QLoRA), so the entire 7B omni-modal pipeline---including
video-frame and, optionally, audio processing---runs on one consumer GPU
(RTX~5090, 32~GB) at a peak memory of 10--21~GB with only about 1.14\% of the
parameters trainable. To attribute any difference to the readout itself
rather than to incidental factors, we run a controlled comparison in which
the backbone, the data, and the LoRA configuration are fixed and only the
readout varies. On CMU-MOSI and CMU-MOSEI the discriminative readout reaches
accuracy on par with recent state-of-the-art systems without task-specific
feature engineering (MOSI: MAE 0.551, Corr 0.888; MOSEI: MAE 0.506, Corr
0.790), and across four random seeds on MOSI it is stable (MAE
$0.570\pm0.014$, Corr $0.877\pm0.009$). Under the controlled comparison the
generative readout---even after equivalent supervised training---more than
doubles the mean absolute error, produces a non-zero rate of unparsable and
out-of-range outputs (about 2.8\% unparsable zero-shot), and is slower at
inference. A modality ablation characterizes the contribution of each input
channel. Our findings indicate that, for continuous MSA, how an LMM is read
out matters as much as how it is trained, and that a discriminative readout
is a more accurate, efficient, and reliable alternative to generative
decoding.

\end{abstract}

\section{Introduction}
Human communication is multimodal: the words people choose, the way they
sound, and their facial expressions jointly convey affect. Multimodal
sentiment analysis (MSA) aims to predict the sentiment expressed in such
signals, often on a continuous scale---for example the interval $[-3,+3]$
used in the widely adopted CMU-MOSI~\cite{mosi} and CMU-MOSEI~\cite{mosei}
benchmarks---with applications in opinion mining, human--computer
interaction, and affective computing~\cite{poria2017review}. Because the
modalities are complementary, with language carrying explicit semantics while
acoustic and visual cues disambiguate tone and emphasis, the fusion of
heterogeneous modalities has long been treated as the central challenge of
MSA.

Early MSA systems relied on hand-crafted or pretrained unimodal features
combined by attention, tensor interactions, or graph
networks~\cite{tfn,lmf,mult}. More recently, large language models (LLMs) and
large multimodal models (LMMs) have reshaped the field: rather than designing
bespoke fusion architectures, a growing body of work adapts a pretrained
(multimodal) LLM to MSA, often with parameter-efficient fine-tuning such as
low-rank adaptation (LoRA)~\cite{lora,qlora}. A defining feature of most such
methods is the readout mechanism---the way a sentiment value is obtained from
the model. The prevailing choice is to treat the model as a text generator
and prompt it to produce the sentiment value as a natural-language string
(e.g.\ \texttt{1.5}), which is then parsed into a
number~\cite{llmsurvey,egmf,loramsa}. We refer to this dominant paradigm as
\emph{generative readout}.

Generative readout is attractive because it reuses the LLM's native decoding
interface and unifies classification and regression under a single
text-generation objective. On closer inspection, though, it is poorly matched
to continuous sentiment regression. It forces an intrinsically continuous
quantity through a discrete, autoregressive token channel, which can limit
precision. Autoregressive decoding is also heavier than a single forward
pass, inflating latency. And free-form generation does not guarantee a valid
number: the model can emit unparsable or out-of-range strings, an instability
that is unacceptable in deployment and has no analogue in a bounded numerical
head. Despite the proliferation of LMM-based MSA, these costs are seldom
quantified; generative readout is adopted largely by default, and whether it
is the right readout for continuous MSA remains underexamined. A further,
practical obstacle is cost: omni-modal LLMs are large, and the perception that
they need multi-GPU clusters discourages their use for MSA.

In this paper we revisit the readout question and argue that, for continuous
MSA, a discriminative readout is preferable. We build on
Qwen2.5-Omni-7B~\cite{qwen25omni}, a native omni-modal LLM whose Thinker
module jointly encodes text, audio, and visual inputs in a shared token
sequence. Rather than letting the model's generative (Talker) head decode a
score, we discard it and read the sentiment directly from the Thinker's
representation: we take the hidden state of the last non-padding token at the
final layer and pass it through a lightweight multilayer perceptron (MLP)
that regresses a scalar sentiment score in one forward pass. To make this
practical and reproducible on commodity hardware, we quantize the backbone to
4-bit NF4 and adapt it with LoRA~\cite{lora,qlora}, training only about
1.14\% of the parameters; with dynamic, pixel-bounded frame sampling, the full
7B pipeline fits on a single 32~GB consumer GPU. To attribute any improvement
to the readout rather than to incidental differences in backbone, data, or
adaptation, we construct a controlled comparison in which all three are fixed
and only the readout---a discriminative head versus generative
decoding---varies. Figure~\ref{fig:arch} illustrates the system.

Across CMU-MOSI and CMU-MOSEI, the discriminative readout matches recent
state-of-the-art systems without specialized feature pipelines, whereas the
generative readout---even when granted the same backbone and additional
supervised training---yields substantially higher error, non-zero unparsable
and out-of-range rates, and slower inference. A modality ablation clarifies
how much each modality contributes and, on CMU-MOSI, reveals a text-dominant
regime that we discuss candidly rather than overstate.

The main contributions are as follows.
\begin{enumerate}
\item \textbf{Readout as a first-class design choice.} We formalize the
readout mechanism as a central, largely overlooked design decision in
LMM-based MSA, and give---to our knowledge---the first controlled,
quantitative comparison of discriminative versus generative readout under an
identical backbone, data, and adaptation budget, jointly measuring accuracy,
inference efficiency, and output reliability.
\item \textbf{A discriminative omni-modal regressor.} We propose a
discriminative hidden-state regression framework on a native omni-modal LLM
(the Qwen2.5-Omni Thinker) that removes the autoregressive generation head
and predicts sentiment from the last-token representation through a
lightweight MLP in one forward pass.
\item \textbf{A reproducible consumer-grade recipe.} We show that 4-bit QLoRA
with dynamic frame sampling lets a 7B omni-modal MSA model train and run on a
single 32~GB consumer GPU at a peak memory of 10--21~GB, lowering the hardware
barrier to LMM-based MSA.
\item \textbf{Competitive results and an honest analysis.} We report accuracy
on par with 2025 state-of-the-art methods on CMU-MOSI and CMU-MOSEI, together
with a modality ablation, a four-seed stability check, and a transparent
discussion of each modality's contribution and of the limitations of
generative readout.
\end{enumerate}

The rest of the paper is organized as follows. Section~\ref{sec:related}
reviews related work. Section~\ref{sec:method} details the discriminative
readout framework and the adaptation recipe. Section~\ref{sec:exp} describes
the setup, the controlled comparison, the main results, the ablation, and an
analysis with limitations. Section~\ref{sec:concl} concludes.

\section{Related Work}\label{sec:related}
\subsection{Multimodal Sentiment Analysis}
MSA extends text-based sentiment analysis by jointly modelling the language,
acoustic, and visual channels of human communication. Two benchmarks have
driven the area: CMU-MOSI~\cite{mosi}, which provides opinion-level continuous
sentiment annotations in $[-3,+3]$ over YouTube monologue clips, and the
larger CMU-MOSEI~\cite{mosei}, which scales the same scheme to tens of
thousands of utterances. Early reviews established the centrality of modality
fusion to affective computing~\cite{poria2017review,baltrusaitis2019survey}.

Much of the field has focused on fusion over pre-extracted unimodal features.
The Tensor Fusion Network (TFN)~\cite{tfn} forms an outer-product tensor to
capture unimodal, bimodal, and trimodal interactions, while Low-rank
Multimodal Fusion (LMF)~\cite{lmf} cuts the resulting cost through low-rank
factorization. The Multimodal Transformer (MulT)~\cite{mult} uses directional
pairwise cross-modal attention to relate unaligned streams without explicit
synchronization. Representation-centric approaches such as MISA~\cite{misa}
split each modality into modality-invariant and modality-specific subspaces
before fusion. With pretrained language models, BERT~\cite{bert} became a
strong textual backbone, and MAG-BERT~\cite{magbert} injected nonverbal cues
into BERT through a multimodal adaptation gate. Self-MM~\cite{selfmm} generated
auxiliary unimodal labels in a self-supervised manner and trained in a
multi-task fashion. Later work explored self-adaptive context
modelling~\cite{selfadaptive} and multi-loss fusion objectives~\cite{mmloss}.
Despite their diversity, these methods share a template: specialized feature
extractors feed a dedicated fusion module whose output is mapped to a sentiment
value by a light task head. They reach strong accuracy but depend on carefully
engineered feature pipelines and bespoke fusion networks.

\subsection{Large Language and Multimodal Models for MSA}
General-purpose LLMs such as GPT-4~\cite{gpt4} and unified multimodal models
have begun to reshape MSA, shifting emphasis from bespoke fusion design toward
adapting powerful pretrained models~\cite{llmsurvey}. Unified formulations such
as UniMSE~\cite{unimse} cast sentiment and emotion within a single text-to-text
interface. Native omni-modal models exemplify the trend:
Qwen2.5-Omni~\cite{qwen25omni}, which we adopt, perceives text, image, audio,
and video in a single end-to-end model through a Thinker--Talker architecture,
where the Thinker performs unified multimodal understanding and the Talker
generates speech.

A defining feature of most LLM-based affective methods is their reliance on
generative readout: the model is prompted to emit the sentiment value as a
string, which is then parsed. Two broad routes have been pursued. Prompt-based
and in-context approaches query a frozen general-purpose model without
parameter updates~\cite{affsurvey}, trading task-specific accuracy for
flexibility; surveys note that, without adaptation, such models still trail
fine-tuned pretrained language models on fine-grained sentiment
tasks~\cite{llmsurvey,affsurvey}. Instruction-tuning approaches instead adapt
an open model on affective instruction data: EmoLLMs~\cite{emollms}, for
example, cast sentiment-strength and emotion-intensity regression as
instruction following and report that prior LLM efforts had largely overlooked
such regression targets in favour of classification, while
Emotion-LLaMA~\cite{emotionllama} extends instruction tuning to multimodal
emotion recognition and reasoning. In the multimodal setting specifically, the
model is commonly adapted with low-rank adaptation for
efficiency~\cite{llmsurvey,egmf,loramsa}, and the sentiment value is still
produced by decoding text.

Across these variants the readout itself is almost invariably generative. This
paradigm reuses the model's native generation interface and unifies
classification and regression under text generation, but it couples a
continuous regression target to discrete autoregressive decoding. That coupling
raises three concerns rarely measured in prior work: an accuracy ceiling from
tokenized numerical outputs, the latency of autoregressive generation relative
to a single forward pass, and the possibility of unparsable or out-of-range
strings. The closest point of contrast is~\cite{egmf}, which shares the
LoRA-adapted-LLM setting but stays within the generative formulation; we
instead remove the generation head entirely and quantify the gap between the
two readout strategies under a controlled setting.

\subsection{Discriminative Readout and Hidden-State Probing}
The alternative to decoding an answer is to read it directly from the model's
internal representation. A growing body of work outside MSA shows that the
hidden states of large pretrained models are highly informative on their own: a
linear or shallow probe trained on final-layer representations recovers
factual, semantic, and safety-relevant properties at accuracy comparable to
much heavier readers~\cite{probehidden,linearsep}. For decoder-only models in
particular, the hidden state of the last token is a natural pooled summary,
because the causal attention mask lets that position attend to the entire
preceding sequence; recent work on text classification exploits exactly this,
attaching a small feed-forward head to the last-token state and finding it
competitive with, and often more stable than, instruction-style generation on
the same backbone~\cite{embedvsinstruct}. Such probes also share a practical
appeal with our design: they require only a single forward pass and a lightweight
head, with no decoding. These observations motivate our approach, which brings
the discriminative, hidden-state readout to multimodal sentiment regression on a
native omni-modal backbone and, unlike prior probing studies that target
classification or interpretability, measures it head-to-head against the
generative readout that dominates MSA.

\subsection{Parameter-Efficient Fine-Tuning}
Fully fine-tuning a billion-parameter model is prohibitive on commodity
hardware. LoRA~\cite{lora} freezes the pretrained weights and learns a small
number of low-rank update matrices, sharply reducing the trainable parameter
count. QLoRA~\cite{qlora} back-propagates through a frozen 4-bit NormalFloat
(NF4) quantized backbone into the LoRA adapters, enabling fine-tuning of very
large models on a single GPU with little loss in quality. These techniques make
it feasible to train and deploy a 7B omni-modal model for MSA on one consumer
GPU, and our recipe is built directly on them.

\subsection{Summary and Positioning}
Classical MSA relies on engineered fusion over unimodal
features~\cite{tfn,lmf,mult,misa,magbert,selfmm}, whereas the emerging
LLM/LMM-based line adapts large pretrained models but predominantly reads
sentiment out by generation~\cite{llmsurvey,egmf,loramsa}. Separately, work on
hidden-state probing shows that a lightweight head on the last-token
representation is a strong and efficient reader in other
domains~\cite{probehidden,linearsep,embedvsinstruct}, yet this discriminative
route has not been brought to bear on continuous multimodal sentiment, nor
compared on equal terms with the prevailing generative readout. Two questions
therefore remain underexplored: whether generative decoding is the right readout
for continuous MSA, and whether such models can be made practical on commodity
hardware. This paper addresses both.

\section{Methodology}\label{sec:method}
\subsection{Problem Formulation and Overview}
We consider continuous multimodal sentiment regression. Each sample is a
triplet $x=(x_t,x_v,x_a)$ comprising a textual transcript $x_t$, a sequence of
video frames $x_v$, and---in the omni-modal configuration---an acoustic stream
$x_a$, with a scalar label $y\in[-3,+3]$. The goal is a mapping
$\hat{y}=F(x)$ that predicts sentiment intensity.

\begin{figure}[t]
\centering
\includegraphics[width=0.95\linewidth]{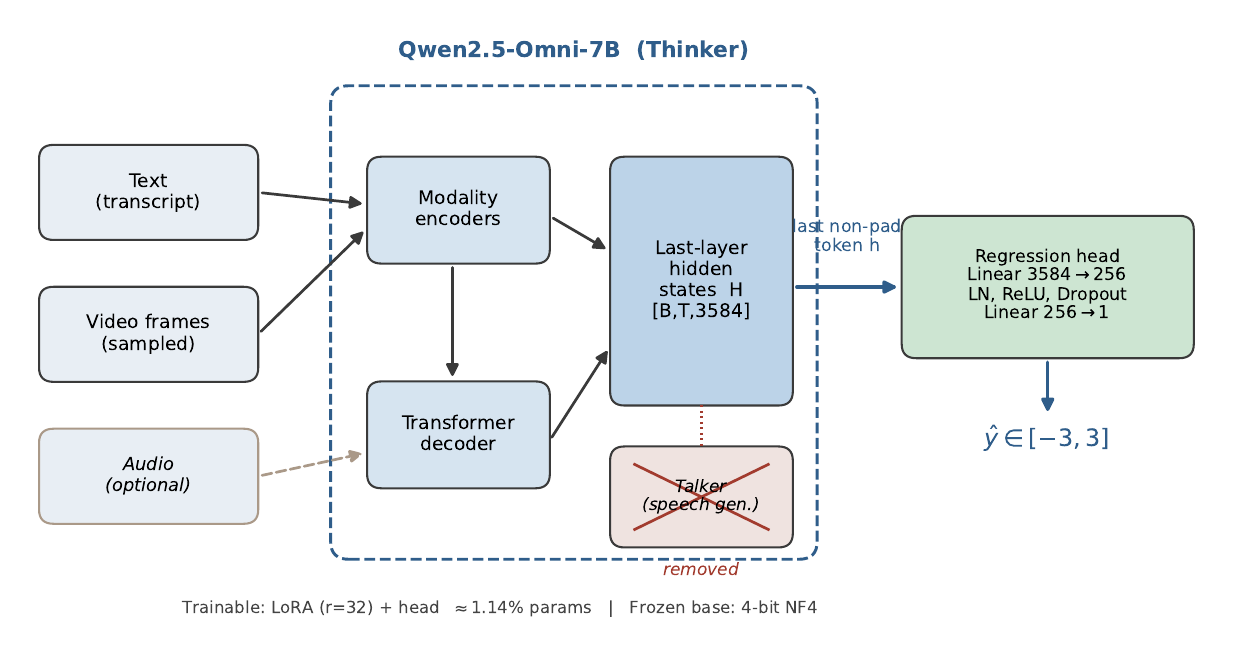}
\caption{Architecture of the proposed discriminative regressor. The
Qwen2.5-Omni-7B Thinker encodes text, video frames, and (optionally) audio
into a unified token sequence. The Talker (speech-generation head) is
discarded. The hidden state of the last non-padding token at the final layer
is passed through a lightweight MLP head to predict $\hat{y}\in[-3,+3]$ in one
forward pass. Trainable components (LoRA adapters and the head, $\approx$103M
parameters, $\approx$1.14\% of the total) are shown in green; the base is
frozen and 4-bit NF4 quantized. The audio path is active only in the
full-modal and ablation configurations.}
\label{fig:arch}
\end{figure}

Unlike the dominant LMM-based approach, which casts this as conditional text
generation and decodes $\hat{y}$ as a string, we factor $F$ into two parts: a
shared omni-modal encoder $f_\theta$ that contextualizes all modalities into a
sequence of hidden states, and a readout that turns those hidden states into a
value. Our central design decision concerns the readout. We propose a
discriminative readout $g_\phi$ that regresses $\hat{y}$ directly from a single
pooled hidden state in one forward pass (Section~\ref{sec:disc}), and contrast
it---under an identical encoder and adaptation budget---with the conventional
generative readout (Section~\ref{sec:gen}). Figure~\ref{fig:arch} gives an
overview: the encoder is the Thinker of Qwen2.5-Omni-7B, adapted with QLoRA,
while the native speech-generation (Talker) head is removed.

\subsection{Omni-Modal Backbone}
We adopt the Thinker of Qwen2.5-Omni-7B~\cite{qwen25omni} as the shared encoder
$f_\theta$. Qwen2.5-Omni is a native omni-modal model: text, images, audio, and
video are processed by modality-specific encoders into tokens that are
concatenated into a single sequence and jointly contextualized by a Transformer
decoder with full self-attention, using a time-aligned multimodal positional
encoding to synchronize the streams. This design is attractive for MSA because
cross-modal interaction is handled internally by the pretrained attention stack,
removing the need for an externally engineered fusion module.

Given a sample $x$, a processor renders the system prompt, the (optionally
present) visual and acoustic tokens, and the transcript into a token sequence of
length $T$, and the Thinker maps it to contextualized hidden states
\begin{equation}
\mathbf{H}=f_\theta(x)=[\mathbf{h}_1,\mathbf{h}_2,\dots,\mathbf{h}_T],
\qquad \mathbf{h}_i\in\mathbb{R}^d,
\end{equation}
where $d=3584$ is the Thinker hidden size. The original model would feed these
to the Talker to autoregressively synthesize speech tokens; we discard the
Talker and operate on $\mathbf{H}$ directly.

\subsection{Discriminative Hidden-State Readout}\label{sec:disc}
The core of our method is a readout that collapses the variable-length sequence
$\mathbf{H}$ into a single scalar without any autoregressive decoding. Given the
causal nature of the Thinker, the last valid token aggregates information from
the whole preceding multimodal context, so we use it as the pooled
representation. Let $\mathbf{m}\in\{0,1\}^T$ be the attention mask marking
non-padding positions. The index of the last non-padding token is
\begin{equation}
\ell=\Big(\textstyle\sum_{i=1}^{T}m_i\Big)-1,
\end{equation}
and the pooled representation is $\mathbf{z}=\mathbf{h}_\ell\in\mathbb{R}^d$.
Selecting $\ell$ via the mask, rather than naively taking the final position
$\mathbf{h}_T$, ensures that right-padding under batched inference never causes
the representation to be read from a padding token---a subtle but important
detail for correctness.

The pooled representation is mapped to a score by a lightweight MLP head
$g_\phi$:
\begin{equation}
\hat{y}=g_\phi(\mathbf{z})=\mathbf{W}_2\,
\mathrm{Dropout}\!\big(\mathrm{ReLU}(\mathrm{LN}(\mathbf{W}_1\mathbf{z}+\mathbf{b}_1))\big)+\mathbf{b}_2,
\end{equation}
where $\mathbf{W}_1\in\mathbb{R}^{256\times d}$,
$\mathbf{W}_2\in\mathbb{R}^{1\times256}$, LN denotes layer normalization, and
dropout (rate 0.2) regularizes the hidden layer. The head has on the order of
$10^6$ parameters and is the only fully task-specific component. The entire
prediction $\hat{y}=g_\phi(f_\theta(x))$ is obtained in one forward pass: no
token-by-token decoding, no sampling, no parsing, so the output is by
construction a bounded real number rather than a string that might fail to
parse.

\subsection{Generative Readout Counterpart}\label{sec:gen}
\begin{figure}[t]
\centering
\includegraphics[width=0.98\linewidth]{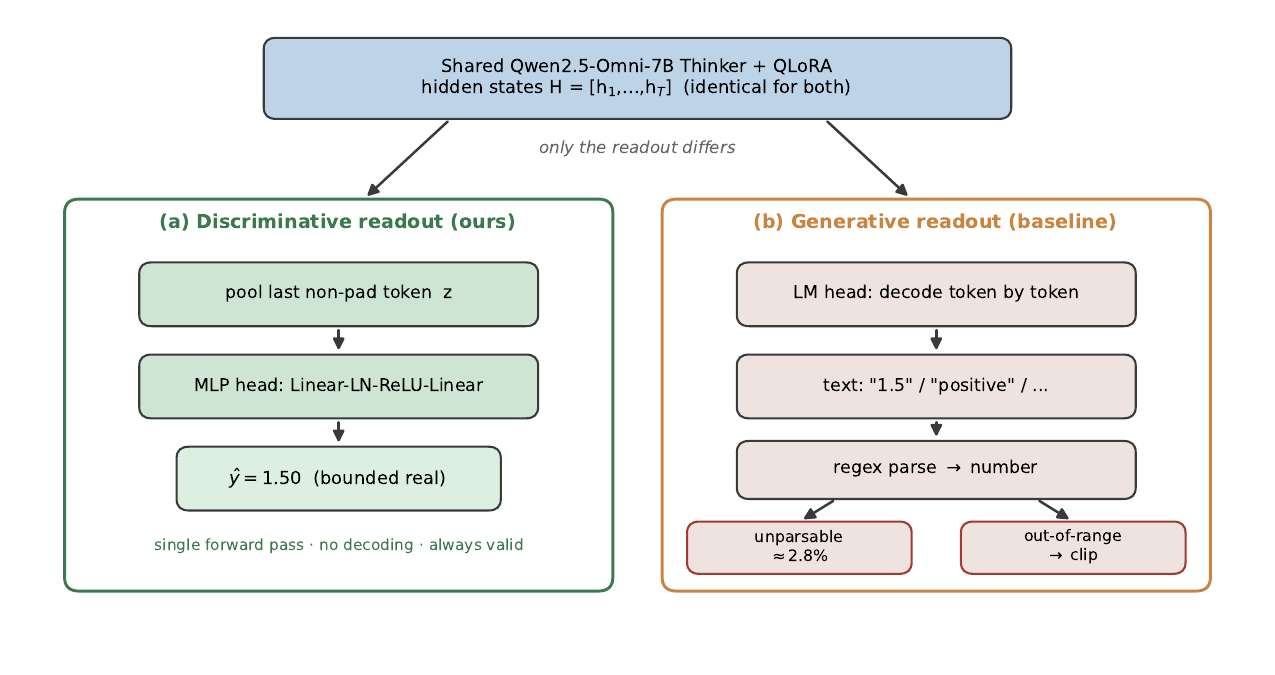}
\caption{The two readout strategies, sharing an identical Qwen2.5-Omni Thinker
backbone and QLoRA adaptation. (a) Our discriminative readout pools the last
non-padding token and projects it to a bounded score in one forward pass, with
a scale-aware regression gradient. (b) The conventional generative readout
autoregressively decodes a text string that must then be parsed, which can be
unparsable or out of range. Isolating the readout as the single controlled
variable enables the like-for-like comparison of Section~\ref{sec:readout}.}
\label{fig:readout}
\end{figure}

To isolate the effect of the readout, we define a generative counterpart that
shares the same backbone $f_\theta$ and the same LoRA adaptation, differing only
in how the sentiment is produced. Instead of the regression head, the model is
asked to generate the value as text. During training the scalar label is
rendered as a numeric string $s(y)$ (e.g.\ $y=1.5\mapsto$ \texttt{1.50}),
appended to the prompt, and the model is optimized with language-modelling
cross-entropy over the target tokens only:
\begin{equation}
\mathcal{L}_{\text{gen}}=-\sum_j \log P_\theta\!\big(s_j\mid x,s_{<j}\big),
\end{equation}
with prompt tokens masked out of the loss. At inference the model greedily
decodes a few tokens, and a numeric value is extracted by a robust parser; if no
valid number is found the output is flagged \emph{unparsable}, and values
outside $[-3,+3]$ are recorded as \emph{out-of-range} (and clipped). These two
failure modes are intrinsic to generative readout and have no analogue in the
discriminative head; we report them as reliability metrics in
Section~\ref{sec:exp}. We evaluate the generative readout both zero-shot
(reusing the backbone without readout-specific training) and after equivalent
supervised training, so the comparison is fair to the generative paradigm.
Figure~\ref{fig:readout} contrasts the two paths: both consume the identical
hidden states of the shared backbone and differ only in the final step---a
single-pass projection to a bounded scalar versus an autoregressive
decode-and-parse pipeline that can fail.

\subsection{Parameter-Efficient Adaptation}
Fully fine-tuning a 7B omni-modal model is infeasible on a single consumer GPU
and would also confound our controlled comparison by changing the backbone. We
therefore freeze the backbone and adapt it with QLoRA~\cite{qlora}. The
pretrained weights are quantized to 4-bit NormalFloat (NF4) with double
quantization and a bfloat16 compute dtype, and low-rank adapters~\cite{lora} are
injected into the attention and feed-forward projections. For a frozen weight
matrix $\mathbf{W}_0\in\mathbb{R}^{d\times k}$, the adapted map is
\begin{equation}
\mathbf{W}_0\mathbf{x}\;\longrightarrow\;
\mathbf{W}_0\mathbf{x}+\tfrac{\alpha}{r}\mathbf{B}\mathbf{A}\mathbf{x},
\qquad \mathbf{B}\in\mathbb{R}^{d\times r},\;\mathbf{A}\in\mathbb{R}^{r\times k},
\end{equation}
with rank $r=32$, scaling $\alpha=32$, and adapter dropout 0.1. Adapters are
applied to the $q,k,v,o,\text{gate},\text{up},\text{down}$ projections. Only the
LoRA matrices $\{\mathbf{A},\mathbf{B}\}$ and the head $g_\phi$ are trainable,
together about 103M parameters, or roughly 1.14\% of the total. Gradient
checkpointing is enabled and the key--value cache is disabled during training to
reduce activation memory; with these measures the full pipeline runs within
10--21~GB of GPU memory. Identical quantization and adapter settings are used for
the generative counterpart, so the two readouts are compared under matched
capacity and memory budgets.

\subsection{Training Objective and Label Normalization}
The discriminative model is trained to minimize the mean absolute error (MAE,
i.e.\ the $L_1$ loss). Because the sentiment distributions of MOSI and especially
MOSEI are concentrated and skewed, a regressor can collapse toward predicting a
near-constant mean. To counteract this we standardize the targets with
training-set statistics. Let $\mu$ and $\sigma$ be the mean and standard
deviation of the training labels; the model is optimized on the normalized
target $\tilde{y}=(y-\mu)/\sigma$:
\begin{equation}
\mathcal{L}_{\text{disc}}=\big|\,g_\phi(f_\theta(x))-\tilde{y}\,\big|,
\end{equation}
and predictions are mapped back at evaluation time via
$\hat{y}=\sigma\,g_\phi(f_\theta(x))+\mu$. We also monitor the standard deviation
of the predictions as a collapse probe: a predicted standard deviation far below
that of the ground truth signals degenerate behaviour.

We use AdamW with a layer-wise schedule that reflects the differing roles of the
components: a smaller rate ($2\times10^{-4}$) for the LoRA adapters, which gently
steer a strong pretrained backbone, and a larger rate ($1\times10^{-3}$) for the
randomly initialized head. A cosine schedule decays the rates, gradient norms are
clipped, and gradients are accumulated over several micro-batches to reach a
larger effective batch under a physical batch size of one. The generative
counterpart uses the analogous schedule on its language-modelling loss; it needs
no label normalization, since it learns to emit numeric strings on the original
scale.

\subsection{Efficient Multimodal Input Pipeline}
A practical bottleneck in video MSA is decoding: naively decoding whole clips is
slow and memory-hungry. We address this on two fronts. First, for each clip we
uniformly sample a fixed number of frames directly by index, avoiding full-clip
decoding, and bound the per-frame resolution between a minimum and maximum pixel
budget. Frame count most strongly affects accuracy, whereas the per-frame pixel
cap is the dominant lever on memory; bounding pixels therefore saves memory at
little cost to accuracy and also speeds up the visual encoder. Second, decoding
runs in a resident worker process guarded by a hard timeout, so occasional
corrupt clips---which can otherwise hang the decoder indefinitely---are skipped
without stalling training, and unreadable clips are filtered out once, ahead of
training, to keep the data distribution stable. A producer thread prefetches and
assembles batches while the GPU computes. These choices are what let the 7B
omni-modal model train end-to-end on a single consumer GPU within a practical
time budget.

\section{Experiments}\label{sec:exp}
Our experiments centre on one question: does the choice of readout matter for
continuous MSA, and by how much? We answer it in three movements. We first
establish that the discriminative readout reaches the accuracy of purpose-built
MSA systems on two benchmarks (Section~\ref{sec:sota}); this is the premise that
makes the rest meaningful, since a readout comparison is interesting only once
the model is competitive. We then turn to the core contribution---a controlled
comparison in which backbone, data, and adaptation are fixed and only the readout
varies (Section~\ref{sec:readout}). Finally we dissect each modality's
contribution through an ablation (Section~\ref{sec:ablation}).

\subsection{Datasets}
We use the two most widely adopted MSA benchmarks. CMU-MOSI~\cite{mosi} contains
2{,}199 opinion utterances from YouTube monologues, each annotated with a
continuous score in $[-3,+3]$. CMU-MOSEI~\cite{mosei} extends the scheme to
roughly 23{,}000 utterances from over 1{,}000 speakers, a larger and more diverse
test bed. We follow the standard speaker-independent train/validation/test splits
released with each dataset. Raw clips and transcripts are fed directly to the
omni-modal model. As preprocessing we denoise the audio track of every clip with
DeepFilterNet~\cite{deepfilternet}, a low-complexity full-band speech enhancement
model; denoising is applied independently per utterance, so no information leaks
across splits. The same denoised clips are used for all experiments on both
datasets. We quantify the effect of this step on CMU-MOSI in
Section~\ref{sec:ablation} (Table~\ref{tab:denoise}).

\subsection{Evaluation Metrics}
Following common practice~\cite{mosei,selfmm,mmsa}, we report regression and
classification metrics. For regression we use MAE (lower is better) and Pearson
correlation (Corr, higher is better). For classification we report seven-class
accuracy (Acc-7), binary accuracy (Acc-2), and the weighted F1. Following our
comparison methods, binary metrics are computed on the non-zero (non-neutral)
subset. As noted in prior work~\cite{almt}, regression and classification metrics
emphasize different aspects---the lowest intensity error need not coincide with
the best classification accuracy---so we report all of them. For the generative
readout we additionally report two reliability metrics intrinsic to that
paradigm: the unparsable rate and the out-of-range rate; both are identically
zero by construction for the discriminative readout.

\subsection{Implementation Details}
All experiments use the Thinker of Qwen2.5-Omni-7B~\cite{qwen25omni} as the
backbone, quantized to 4-bit NF4 with a bfloat16 compute dtype and adapted with
LoRA of rank $r=32$, scaling $\alpha=32$, and dropout 0.1 on the
$\{q,k,v,o,\text{gate},\text{up},\text{down}\}$ projections. We optimize with
AdamW under a layer-wise schedule (LoRA $2\times10^{-4}$, head $1\times10^{-3}$)
with cosine annealing, gradient clipping, and a physical batch size of one with
gradient accumulation for a larger effective batch. From each clip we uniformly
sample up to 16 frames with a bounded per-frame pixel budget. All runs use a
single NVIDIA RTX~5090 (32~GB); the complete pipeline, including video decoding,
operates within a peak memory of 10--21~GB, confirming that a 7B omni-modal MSA
model is trainable and deployable on one consumer GPU.

\subsection{Comparison with the State of the Art}\label{sec:sota}
We first verify that, with the discriminative readout, a single-GPU omni-modal
LLM matches purpose-built MSA systems. Table~\ref{tab:mosei} compares our model
against a broad range of methods on CMU-MOSEI, spanning four generations of MSA
research: classical tensor- and graph-based fusion (TFN~\cite{tfn},
LMF~\cite{lmf}, MFN~\cite{mfn}), cross-modal Transformers (MulT~\cite{mult}),
BERT-based fusion (MISA~\cite{misa}, MAG-BERT~\cite{magbert}), self-supervised
and mutual-information methods (Self-MM~\cite{selfmm}, MMIM~\cite{mmim}),
contrastive and distillation-based learning (ConFEDE~\cite{confede},
DMD~\cite{dmd}), the language-guided Transformer ALMT~\cite{almt}, and recent
2025--2026 systems including the state-space model MSAmba~\cite{msamba}, the
modality-enhanced fusion model MEMMI~\cite{memmi}, and the hierarchical-alignment
model DecAlign~\cite{decalign}.

\begin{table}[t]
\centering
\caption{Comparison with state-of-the-art methods on CMU-MOSEI. Higher is better
except for MAE; ``--'' denotes a value not reported in the source. Acc-2 and F1
use the non-zero (negative/positive) convention. Best results in bold. Our model
uses the full-modal discriminative readout under the best training configuration.
Classical baselines are taken from the unified MMSA benchmark~\cite{mmsa}; more
recent methods are quoted from their original papers.}
\label{tab:mosei}
\small
\begin{tabular}{lccccc}
\toprule
Method & MAE\,\dn & Corr\,\up & Acc-7\,\up & Acc-2\,\up & F1\,\up \\
\midrule
TFN~\cite{tfn}        & 0.593 & 0.700 & 50.2 & 82.5 & 82.1 \\
LMF~\cite{lmf}        & 0.623 & 0.677 & 48.0 & 82.0 & 82.1 \\
MFN~\cite{mfn}        & 0.568 & 0.717 & 51.1 & 84.0 & 83.9 \\
MulT~\cite{mult}      & 0.580 & 0.703 & 51.8 & 82.5 & 82.3 \\
MISA~\cite{misa}      & 0.555 & 0.756 & 52.2 & 85.5 & 85.3 \\
MAG-BERT~\cite{magbert}& 0.539 & 0.753 & 52.7 & 85.2 & 85.1 \\
Self-MM~\cite{selfmm} & 0.530 & 0.765 & 53.6 & 85.2 & 85.3 \\
MMIM~\cite{mmim}      & 0.526 & 0.772 & 54.2 & 85.9 & 85.3 \\
ConFEDE~\cite{confede}& 0.522 & 0.780 & 54.9 & 85.8 & 85.8 \\
DMD~\cite{dmd}        & 0.532 & 0.766 & 54.0 & 86.0 & 85.9 \\
ALMT~\cite{almt}      & 0.526 & 0.779 & 53.7 & 86.4 & 86.4 \\
MSAmba~\cite{msamba}  & 0.521 & 0.781 & 54.4 & 86.5 & 86.4 \\
MEMMI~\cite{memmi}    & 0.526 & 0.779 & 54.2 & 86.0 & 86.0 \\
DecAlign~\cite{decalign}& 0.543 & 0.768 & \textbf{55.0} & 86.5 & 86.1 \\
\midrule
Ours (discriminative) & \textbf{0.506} & \textbf{0.790} & \textbf{55.0} & \textbf{87.1} & \textbf{87.0} \\
\bottomrule
\end{tabular}
\end{table}

The proposed model attains MAE 0.506 and correlation 0.790 on CMU-MOSEI, on par
with---and on these two metrics marginally better than---the strongest recent
comparators, while using neither task-specific feature extractors nor a
hand-designed fusion network. Two points deserve emphasis. Every competing method
in Table~\ref{tab:mosei} relies on externally extracted unimodal features (e.g.\
COVAREP for audio, OpenFace for vision) feeding a bespoke fusion module; our model
consumes raw frames and text through a single pretrained omni-modal backbone, so
reaching the same accuracy with a far simpler pipeline is itself a meaningful
result. And, consistent with the wider literature, no single method dominates
every metric: classification- and regression-oriented systems trade places across
columns~\cite{almt}. We therefore make no claim of universally outperforming the
field; the narrower, robust result we rely on is that the discriminative readout
is competitive with the state of the art, which is the premise the rest of this
section builds upon.

Table~\ref{tab:mosi} reports the same comparison on the smaller CMU-MOSI
benchmark, on which the bulk of our analysis is conducted. Here our model attains
MAE 0.551 and correlation 0.888, again competitive with the strongest recent
methods; the correlation in particular is among the highest reported, indicating
that the discriminative readout captures the relative ordering of sentiment
intensities well even on this smaller corpus. As is widely observed, absolute
scores on CMU-MOSI are noisier than on CMU-MOSEI because of its limited size, so
we treat the two benchmarks jointly rather than over-interpreting any single
metric on either one.

\begin{table}[t]
\centering
\caption{Comparison with state-of-the-art methods on CMU-MOSI. Higher is better
except for MAE. Acc-2 and F1 use the non-zero (negative/positive) convention. Best
results in bold. Our model uses the full-modal discriminative readout under the
best training configuration. Classical baselines are taken from the unified MMSA
benchmark~\cite{mmsa}; more recent methods are quoted from their original papers.}
\label{tab:mosi}
\small
\begin{tabular}{lccccc}
\toprule
Method & MAE\,\dn & Corr\,\up & Acc-7\,\up & Acc-2\,\up & F1\,\up \\
\midrule
TFN~\cite{tfn}        & 0.947 & 0.673 & 34.5 & 79.1 & 79.1 \\
LMF~\cite{lmf}        & 0.950 & 0.651 & 33.8 & 79.2 & 79.2 \\
MulT~\cite{mult}      & 0.880 & 0.702 & 36.9 & 81.0 & 81.0 \\
MISA~\cite{misa}      & 0.777 & 0.778 & 41.4 & 83.5 & 83.6 \\
MAG-BERT~\cite{magbert}& 0.731 & 0.789 & 43.6 & 84.3 & 84.3 \\
Self-MM~\cite{selfmm} & 0.713 & 0.798 & 46.7 & 85.0 & 84.9 \\
MMIM~\cite{mmim}      & 0.700 & 0.800 & 46.7 & 85.1 & 85.0 \\
ConFEDE~\cite{confede}& 0.742 & 0.784 & 42.3 & 85.5 & 85.5 \\
DMD~\cite{dmd}        & 0.723 & 0.794 & 45.6 & 85.7 & 85.6 \\
ALMT~\cite{almt}      & 0.683 & 0.805 & 47.9 & 85.6 & 85.6 \\
MSAmba~\cite{msamba}  & 0.681 & 0.806 & 47.0 & 86.0 & 86.0 \\
\midrule
Ours (discriminative) & \textbf{0.551} & \textbf{0.888} & \textbf{52.9} & \textbf{89.5} & \textbf{89.5} \\
\bottomrule
\end{tabular}
\end{table}

\paragraph{Stability across seeds.}
Because the CMU-MOSI training set is small, we check that the result is not an
artefact of a single lucky run. We retrain the full-modal discriminative model
under four random seeds and report the mean and standard deviation
(Table~\ref{tab:seeds}). The spread is narrow---MAE $0.570\pm0.014$ and Corr
$0.877\pm0.009$---and every seed converges to a similar point, with no collapsed
or divergent run. The single-run figures in Table~\ref{tab:mosi} (MAE 0.551, Corr
0.888) correspond to the best of these seeds; we report the seed-averaged numbers
here so the result is not read as seed-dependent.

\begin{table}[t]
\centering
\caption{Stability of the discriminative readout on CMU-MOSI across four random
seeds. Acc-2 and F1 use the non-zero convention. The best single run is the one
reported in Table~\ref{tab:mosi}.}
\label{tab:seeds}
\small
\begin{tabular}{lccccc}
\toprule
Seed & MAE\,\dn & Corr\,\up & Acc-7\,\up & Acc-2\,\up & F1\,\up \\
\midrule
A & 0.551 & 0.888 & 52.9 & 89.5 & 89.5 \\
B & 0.569 & 0.865 & 52.3 & 88.8 & 88.7 \\
C & 0.576 & 0.879 & 51.0 & 88.4 & 88.5 \\
D & 0.584 & 0.875 & 53.0 & 89.8 & 89.8 \\
\midrule
Mean $\pm$ std & $0.570\pm0.014$ & $0.877\pm0.009$ & $52.3\pm0.9$ & $89.1\pm0.7$ & $89.1\pm0.7$ \\
\bottomrule
\end{tabular}
\end{table}

\subsection{Discriminative versus Generative Readout}\label{sec:readout}
We now address the core question. To isolate the effect of the readout, this
comparison uses an identical backbone, identical data, and an identical LoRA
configuration; the only variable is how the sentiment is produced---a
discriminative head versus generative text decoding
(Sections~\ref{sec:disc}--\ref{sec:gen}). Because this controlled protocol fixes
the training budget at a common number of epochs for every variant, the
discriminative numbers here are not identical to the best-tuned results of
Tables~\ref{tab:mosei}--\ref{tab:mosi}; the two settings answer different
questions---best attainable accuracy versus a like-for-like readout
comparison---and should be read accordingly. We evaluate the generative readout in
two regimes: zero-shot, reusing the backbone without readout-specific training,
and trained, after equivalent supervised fine-tuning on the numeric-string target,
so the comparison is fair to the generative paradigm rather than a straw man.

\begin{figure}[t]
\centering
\includegraphics[width=0.98\linewidth]{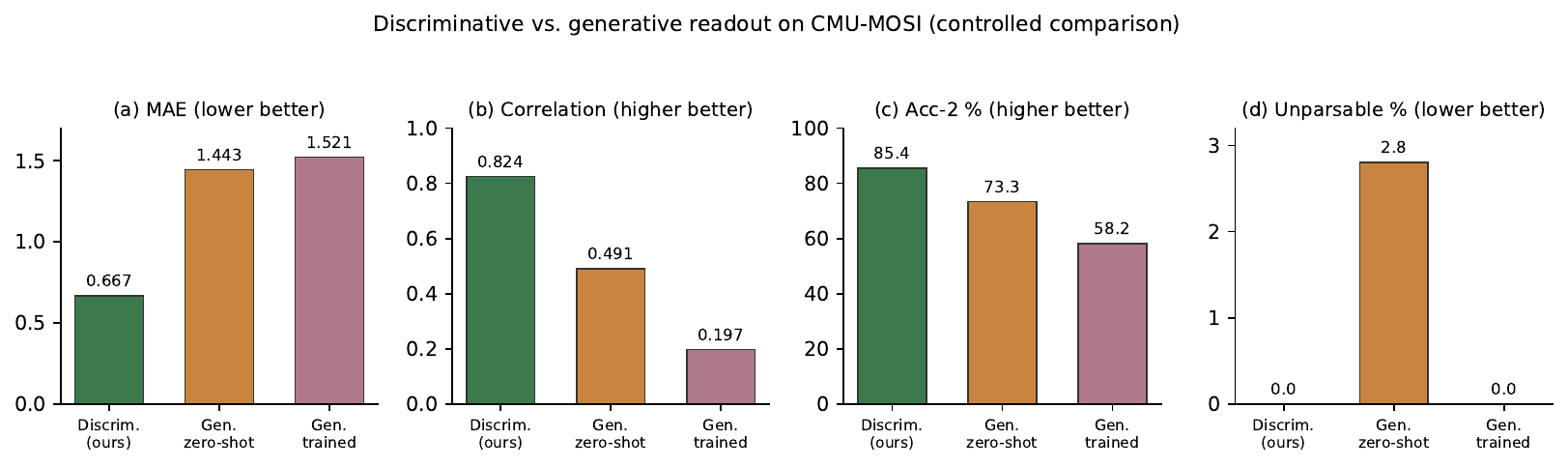}
\caption{Visual summary of the controlled readout comparison on CMU-MOSI
(Table~\ref{tab:readout}). Across accuracy (MAE, correlation, Acc-2) and output
reliability (unparsable rate), the discriminative readout consistently and
substantially outperforms both the zero-shot and the trained generative readout.
All axes start at zero. Additional training does not help the generative
readout---its correlation degrades---while the discriminative head incurs no
unparsable outputs by construction.}
\label{fig:bars}
\end{figure}

Table~\ref{tab:readout} reports the comparison and Figure~\ref{fig:bars}
visualizes it; three findings stand out.
\emph{(i) Accuracy.} The discriminative readout is far more accurate: on CMU-MOSI
it more than halves the MAE relative to the generative readout (0.667 vs.\ 1.443
zero-shot) and roughly doubles the correlation, with a similar gap on CMU-MOSEI.
\emph{(ii) Training does not rescue generation.} Supervising the generative
readout does not close the gap---its correlation in fact degrades
($0.491\to0.197$ on MOSI). This is diagnostic rather than incidental: optimizing a
model to emit a two-decimal number string is a weak, ill-posed signal for a
continuous target (the loss is dominated by token-level formatting rather than by
sentiment magnitude), whereas the discriminative head receives a direct,
scale-aware regression gradient. \emph{(iii) Reliability and speed.} The
discriminative readout is failure-free by construction, while the generative
readout produces about 2.8\% unparsable outputs and a non-zero out-of-range
rate---behaviour that is unacceptable in deployment and absent from a bounded
numerical head. The discriminative readout is also faster per sample, needing a
single forward pass rather than autoregressive decoding; peak memory is essentially
identical, as expected given the shared backbone, so the efficiency advantage is
one of speed and reliability rather than memory.

\begin{table}[t]
\centering
\caption{Controlled comparison of readout strategies under an identical backbone,
data, and LoRA configuration (common training budget). Unparsable and OOB
(out-of-range) are reliability metrics intrinsic to generative decoding; they are
zero by construction for the discriminative head. ``--'' denotes a configuration
we did not run. Best values in bold.}
\label{tab:readout}
\small
\begin{tabular}{llccc}
\toprule
& & Discriminative & \multicolumn{2}{c}{Generative} \\
\cmidrule(lr){4-5}
Dataset & Metric & (ours) & zero-shot & trained \\
\midrule
\multirow{3}{*}{MOSI}
 & MAE\,\dn  & \textbf{0.667} & 1.443 & 1.521 \\
 & Corr\,\up & \textbf{0.824} & 0.491 & 0.197 \\
 & Acc-2\,\up& \textbf{85.4}  & 73.3  & 58.2 \\
\midrule
\multirow{2}{*}{MOSEI}
 & MAE\,\dn  & \textbf{0.521} & 1.431 & -- \\
 & Corr\,\up & \textbf{0.790} & 0.473 & -- \\
\midrule
\multicolumn{2}{l}{Unparsable\,\dn}      & \textbf{0.0\%} & 2.8\% & 0.0\% \\
\multicolumn{2}{l}{OOB\,\dn}             & \textbf{0.0\%} & 0.05\% & 0.0\% \\
\multicolumn{2}{l}{Peak mem.\ (GB)\,\dn} & 10.78 & 10.78 & -- \\
\multicolumn{2}{l}{Inf.\ time (s/sample)\,\dn} & \textbf{1.14} & 1.47 & -- \\
\bottomrule
\end{tabular}
\end{table}

Taken together, these results answer the paper's central question. The readout is
not an incidental implementation detail but a decisive design factor: under matched
backbone, data, and capacity, the discriminative choice is better on accuracy,
stability, and speed at the same time. This is, to our knowledge, the first time
the cost of generative readout for continuous MSA has been quantified under
controlled conditions.

Why the readout matters so much has a single underlying cause: the mismatch
between a continuous target and a discrete generation channel. A regression head
receives a smooth, scale-aware gradient that directly penalizes the magnitude of
its error, whereas a generative model is trained to maximize the likelihood of a
particular numeric string, a signal largely indifferent to whether 1.4 is closer
to the truth than 1.9. This also explains the otherwise puzzling observation that
training the generative readout did not help, and even hurt its correlation:
additional supervision sharpened the model's formatting of the output string
without improving its estimate of sentiment intensity. The discriminative readout
avoids this mismatch entirely, which is why it is at once more accurate, more
reliable, and faster. Beyond accuracy this carries a practical message: a 7B
native omni-modal model, adapted with 4-bit QLoRA and dynamic frame sampling, can
be trained and deployed for MSA entirely on a single 32~GB consumer GPU at a peak
memory of 10--21~GB and with a failure-free, single-pass inference interface,
lowering the hardware barrier that has discouraged the use of omni-modal LLMs for
MSA.

\subsection{Modality Ablation}\label{sec:ablation}
To understand which input channels drive the prediction, we ablate the modalities
fed to the discriminative model on CMU-MOSI, training each configuration under the
same controlled budget so the rows are directly comparable. Table~\ref{tab:abl}
reports text-only, text+video, and the full configuration. The picture is nuanced
rather than uniformly additive. On the regression metrics (MAE and correlation),
text-only is strongest, and adding the visual and acoustic channels slightly
increases the error; on the classification metrics (Acc-2 and F1), text+video is
best. This indicates that CMU-MOSI is a strongly text-dominant dataset, on which
the nonverbal channels carry limited complementary signal for fine-grained
intensity regression and can even introduce noise---a tendency also reflected in
the text-centric design of several recent methods~\cite{confede,tetfn}. We read
this as a property of the dataset rather than of our method: CMU-MOSI consists of
short opinion clips in which the spoken content is highly informative while the
visual and acoustic cues are comparatively weak. We report the text-dominant
effect candidly rather than overstating the benefit of multimodality on this
benchmark. Crucially, this dataset-specific observation does not weaken the paper's
central claim, which concerns the readout mechanism and holds across both
benchmarks regardless of the modality mix.

\begin{table}[t]
\centering
\caption{Modality ablation on CMU-MOSI with the discriminative readout, under a
common training budget. Acc-2 and F1 use the non-zero convention. Best values in
bold.}
\label{tab:abl}
\small
\begin{tabular}{lccccc}
\toprule
Configuration & MAE\,\dn & Corr\,\up & Acc-2\,\up & F1\,\up & Acc-7\,\up \\
\midrule
Text only      & \textbf{0.552} & \textbf{0.883} & 87.3 & 87.3 & -- \\
Text + Video   & 0.631 & 0.858 & \textbf{88.1} & \textbf{88.0} & 46.8 \\
Full (T+V+A)   & 0.667 & 0.824 & 85.4 & 85.4 & -- \\
\bottomrule
\end{tabular}
\end{table}

\paragraph{Effect of audio denoising.}
Since our pipeline denoises the audio track of every clip
(Section~\ref{sec:exp}), we isolate the contribution of this step on CMU-MOSI by
training the full-modal discriminative model on the original and the denoised clips
under an otherwise identical configuration. Table~\ref{tab:denoise} reports the
result. Denoising improves the regression metrics---reducing MAE from 0.598 to
0.551 and raising correlation from 0.878 to 0.888---while binary accuracy is
essentially unchanged (89.63\% $\to$ 89.52\%). Cleaning the acoustic channel
sharpens the fine-grained intensity estimate without altering the coarse polarity
decision, consistent with the audio modality carrying intensity-related prosodic
cues. We apply the same denoising to CMU-MOSEI for consistency but, owing to
compute constraints, did not repeat this ablation there; we note this in
Section~\ref{sec:limit}.

\begin{table}[t]
\centering
\caption{Effect of audio denoising with DeepFilterNet~\cite{deepfilternet} on
CMU-MOSI, using the full-modal discriminative model under an identical
configuration. The denoised row corresponds to the best run in
Table~\ref{tab:mosi}. Best values in bold.}
\label{tab:denoise}
\small
\begin{tabular}{lccc}
\toprule
Configuration & MAE\,\dn & Corr\,\up & Acc-2\,\up \\
\midrule
Original audio & 0.598 & 0.878 & \textbf{89.63} \\
Denoised audio & \textbf{0.551} & \textbf{0.888} & 89.52 \\
\bottomrule
\end{tabular}
\end{table}

\subsection{Representation-Space Analysis}
\begin{figure}[t]
\centering
\includegraphics[width=0.92\linewidth]{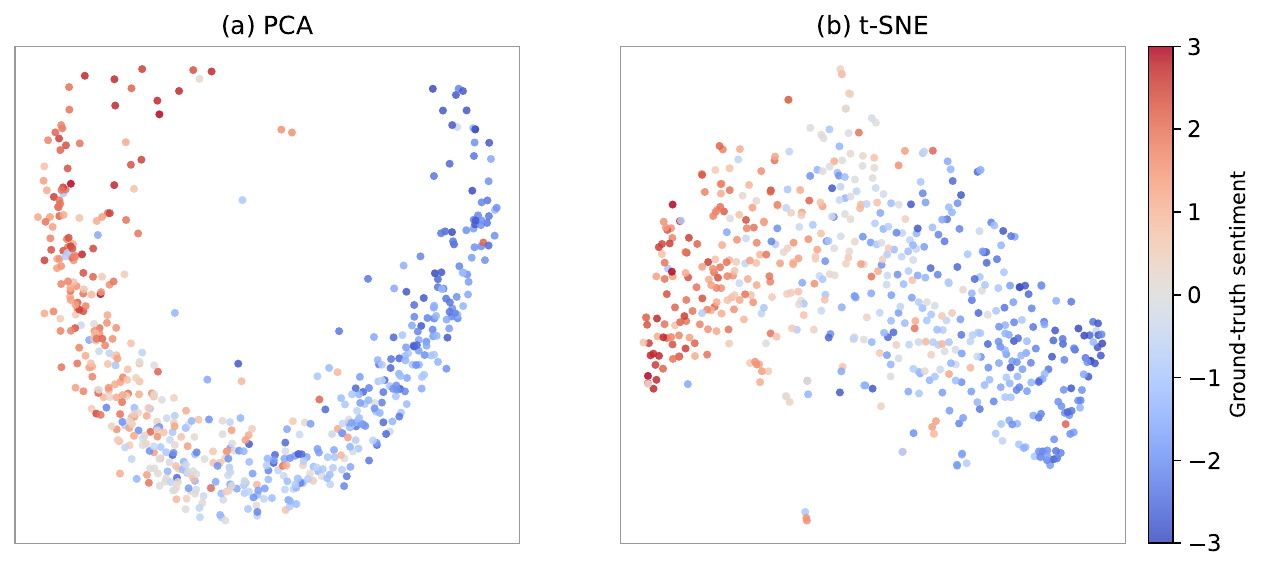}
\caption{Two-dimensional (a) PCA and (b) t-SNE projections of the last-token
hidden states of the discriminative model on the CMU-MOSI test set (600 clips),
coloured by ground-truth sentiment ($-3$ to $+3$). Both projections show a smooth
negative-to-positive gradient, with polarity already well separated under the
linear PCA projection. The $k$NN label-smoothness statistic (0.950 vs.\ 1.896 for
permuted labels) confirms quantitatively that the representation is organized by
sentiment intensity.}
\label{fig:hidden}
\end{figure}

The quantitative gap in favour of the discriminative readout
(Section~\ref{sec:readout}) raises a natural question: is the last-token hidden
state of the omni-modal backbone actually organized according to sentiment
intensity, such that a simple readout suffices? To examine this directly we
extract, for every test utterance, the final-layer hidden state of the last
non-padding token---exactly the vector consumed by our regression head---and
visualize it in two dimensions with PCA and t-SNE~\cite{tsne}, colouring each point
by its ground-truth sentiment. Figure~\ref{fig:hidden} shows the result; both
projections reveal a smooth gradient from negative to positive sentiment, and the
linear PCA projection already separates polarity cleanly, which indicates that the
sentiment direction is largely linear in the hidden space.

To avoid relying on visual impression alone, we report a $k$-nearest-neighbour
label-smoothness statistic: for each sample we average the absolute label
difference to its $k=10$ nearest neighbours in the hidden space and compare it
against the same quantity computed on randomly permuted labels. A neighbourhood
difference well below the permuted baseline indicates that the representation is
organized along the sentiment-intensity continuum. On CMU-MOSI we measure a
neighbourhood difference of 0.950 against a permuted baseline of 1.896, a 49.9\%
reduction, confirming that nearby hidden states share similar sentiment; the value
is stable across neighbourhood sizes (0.49--0.51 for $k=5,10,20$). This gives a
representational explanation for why a lightweight discriminative head is
sufficient: the pretrained omni-modal Thinker already arranges multimodal context
into a sentiment-graded geometry, so the readout need only project it onto a
scalar, whereas generative decoding must re-encode that geometry into discrete
tokens. The few points that lie far from neighbours of the same colour are the
boundary and hard cases present in any annotated set.

\subsection{Limitations}\label{sec:limit}
We state the limitations plainly. First, our seed-averaged stability check
(Table~\ref{tab:seeds}) was run on CMU-MOSI; we report a single converged run on
CMU-MOSEI owing to its larger size and the associated compute cost, and a
seed-averaged evaluation on MOSEI would further strengthen the claims. Second, the
modality ablation and the audio-denoising ablation were conducted on CMU-MOSI
only; the corresponding ablations on CMU-MOSEI were not performed owing to compute
constraints, and we mark this explicitly rather than extrapolating. Third, the
comparisons in Tables~\ref{tab:mosei}--\ref{tab:mosi} draw classical-baseline
numbers from the unified MMSA benchmark~\cite{mmsa} and recent numbers from the
respective papers, which differ in feature extractors and evaluation conventions;
we adopt a consistent non-zero convention here, but a fully unified re-evaluation
of every method under one framework remains future work. Finally, our findings
concern continuous sentiment regression on English benchmarks; whether the
discriminative-readout advantage transfers to categorical emotion recognition and
to other languages is an open question.

\section{Conclusion}\label{sec:concl}
We revisited a design choice that most LLM-based multimodal sentiment analysis
takes for granted---how the sentiment value is read out of the model---and argued
that, for continuous MSA, a discriminative readout is preferable to the prevailing
generative one. Building on the Thinker of the native omni-modal model
Qwen2.5-Omni-7B, we discarded the autoregressive generation head and regressed the
sentiment directly from the last-token hidden state through a lightweight MLP, in
one forward pass, adapting the backbone with 4-bit QLoRA so the entire 7B pipeline
runs on a single consumer GPU. Under a controlled protocol that fixes the
backbone, data, and adaptation and varies only the readout, the discriminative
formulation proved at once more accurate (more than halving the MAE on CMU-MOSI),
more reliable (zero unparsable or out-of-range outputs versus a non-zero rate for
generation), and faster, while matching recent state-of-the-art systems on both
CMU-MOSI and CMU-MOSEI without task-specific feature engineering. A four-seed
check confirmed the stability of the result, and a modality ablation characterized
the text-dominant nature of CMU-MOSI, which we discussed transparently. For
continuous MSA, how an omni-modal LLM is read out is as consequential as how it is
trained. Future work includes seed-averaged evaluation on CMU-MOSEI, modality
ablation on CMU-MOSEI, a unified re-evaluation of baselines, and an extension of
the discriminative-readout analysis to categorical emotion recognition and
multilingual settings.

\end{document}